\documentclass[ALICE,manyauthors]{cernphprep}

\usepackage[comma,square,numbers,sort&compress]{natbib}
\usepackage{hyperref}

\usepackage{lineno}
\usepackage{xspace}

\usepackage{multirow}
\usepackage{bm}
\usepackage{diagbox}
\usepackage{makecell}
\usepackage{color}
\usepackage{graphicx}
\usepackage[T1]{fontenc}

\def \psitwoep {\Psi_{\rm 2}^{\rm EP}}
\def \psitwo {\Psi_{\rm 2}}
\def \pzstwo {P_{\rm z,\,{\rm s2}}}

\newcommand{\pt}{p_{\rm T}}

\newcommand{\sNN}{\ensuremath{\sqrt{s_{_{\rm NN}}}}\xspace}

\newcommand{ \la }{\langle}
\newcommand{ \ra }{\rangle}

\newcommand{\mean}[1]{\la #1 \ra}

\newcommand{ \lam }{{$\Lambda$}\xspace}

\definecolor{dgreen}{cmyk}{1.,0.,1.,0.2} 
\definecolor{orange}{cmyk}{0.,0.353,1.,0.} 

\begin{document}

\begin{titlepage}
\PHyear{2021}       
\PHnumber{148}      
\PHdate{21 July}  

\title{Polarization of $\Lambda$ and $\overline{\Lambda}$ hyperons along the beam
direction in Pb--Pb collisions at $\sNN=5.02$~TeV}

\ShortTitle{Polarization of $\Lambda$ ($\overline{\Lambda}$) hyperons in Pb--Pb collisions}   

\Collaboration{ALICE Collaboration\thanks{See Appendix~\ref{app:collab} for the list of collaboration members}}
\ShortAuthor{ALICE Collaboration} 

\begin{abstract}
The polarization of the $\Lambda$ and $\overline\Lambda$ hyperons
along the beam ($z$) direction, $P_{\rm z}$, has been measured in
Pb--Pb collisions at $\sNN=5.02$~TeV recorded with ALICE at the Large
Hadron Collider (LHC).  The main contribution to $P_{\rm z}$ comes
from elliptic flow induced vorticity and can be characterized by the
second Fourier sine coefficient $\pzstwo = \mean{P_{\rm z}
  \sin(2\varphi - 2 \psitwo)}$, where $\varphi$ is the hyperon
azimuthal emission angle, and $\psitwo$ is the elliptic flow plane
angle. We report the measurement of $\pzstwo$ for different collision
centralities, and in the 30--50\% centrality interval as a function of
the hyperon transverse momentum and rapidity. The $\pzstwo$ is
positive similarly as measured by the STAR Collaboration in Au--Au
collisions at $\sNN = 200$~GeV, with somewhat smaller amplitude in the
semi-central collisions. This is the first experimental evidence of a
non-zero hyperon $P_{\rm z}$ in Pb--Pb collisions at the LHC. 
The comparison of the measured $\pzstwo$ with the hydrodynamic model
calculations shows sensitivity to the competing contributions from
thermal and the recently found shear induced vorticity, as well as to
whether the polarization is acquired at the quark--gluon plasma or the hadronic phase.

\end{abstract}
\end{titlepage}
\setcounter{page}{2}


The system created in high-energy nuclear collisions behaves almost
like an ideal fluid~\cite{Braun-Munzinger:2015hba}. Its evolution is
characterized by non-trivial velocity and vorticity fields, resulting
in the polarization of the produced particles. In particular, the
shear in the initial velocity distributions of the participants in
off-center nuclear collisions leads to a non-zero vorticity component
and a net particle polarization along the orbital momentum of the
colliding nuclei, a phenomenon termed as global
polarization~\cite{Liang:2004ph,Voloshin:2004ha,Becattini:2015ska}.
Recent measurements at RHIC show a significant global polarization of
$\Lambda$ and $\overline\Lambda$ hyperons in Au--Au collisions at
$\sqrt{s_{\rm NN}} = $ 7.7 -- 200 GeV with the polarization magnitude
of a few to a fraction of a percent, monotonically decreasing with
increasing $\sqrt{s_{\rm NN}}$~\cite{STAR:2017ckg, Adam:2018ivw}. The
global hyperon polarization measured by the ALICE Collaboration in
Pb--Pb collisions at $\sqrt{s_{\rm NN}} = $ 2.76 and
5.02~TeV~\cite{Acharya:2019ryw} was found to be at the per mil level,
consistent with zero within experimental uncertainties. The ALICE
measurements are also consistent with hydrodynamical model
calculations for the LHC energies and empirical estimates based on the
collision energy dependence of the directed flow due to the tilted
source~\cite{Becattini:2015ska, Voloshin:2017kqp, Abelev:2013cva}. The
decrease in the global polarization at midrapidity with collision
energy is usually attributed to a decreasing role of the baryon
stopping~\cite{Sorge:1991pg} in the initial velocity distributions.

In addition to the vorticity due to the orbital angular momentum of
the entire system, other physics processes, such as anisotropic flow,
jet energy deposition, and deviation from longitudinal boost invariance of
the transverse velocity fields, generate vorticity~\cite{Betz:2007kg,
  Voloshin:2017kqp, Becattini:2017gcx, Pang:2016igs, Xia:2018tes,
  Sun:2018bjl} along different directions depending on the location of
the fluid elements in the created system. It was
predicted that in non-central nucleus-nucleus collisions, the strong
elliptic flow would generate a non-zero vorticity component along the
beam axis ($z$)~\cite{Voloshin:2017kqp,Becattini:2017gcx}. The
vorticity and the corresponding polarization exhibits a quadrupole
structure in the transverse plane. This polarization, characterized by the second harmonic sine component in the Fourier decomposition of the polarization along the beam axis ($P_{\rm z}$) as a function of the particle azimuthal angle ($\varphi$) relative to the elliptic flow plane $\psitwo$, is evaluated as
\begin{equation}
\pzstwo = \mean{{P_{\rm z} \sin(2\varphi - 2 \psitwo)}}. 
\label{Eq11}   
\end{equation}
The sign of $\pzstwo$ determines the
phase of the $P_{\rm z}$ modulation in azimuth relative to the
elliptic flow plane.

The \lam and $\overline\Lambda$ polarization along the beam direction
was measured by the STAR Collaboration in Au--Au collisions at
$\sqrt{s_{\rm NN}} = 200$~GeV~\cite{Adam:2019srw} and compared with the
hydrodynamic~\cite{Becattini:2017gcx}, transport
(AMPT)~\cite{Xia:2018tes,Fu:2020oxj}, and Blast-Wave
(BW)~\cite{Voloshin:2017kqp,Adam:2019srw} model calculations. The
measured $\pzstwo$ was found to be about 5 times smaller in magnitude
and of opposite sign compared to the hydrodynamic and AMPT model
predictions. However, the BW model, tuned to spectra, elliptic flow,
and azimuthally differential femtoscopic measurements, describes the magnitude and the sign of $\pzstwo$.
Most model calculations estimate the particle polarization from the
thermal vorticity~\cite{Becattini:2017gcx, Fu:2020oxj} at the
freeze-out surface assuming local thermodynamic equilibrium of the
spin degrees of freedom.  Unlike hydrodynamic and AMPT calculations,
the BW model~\cite{Voloshin:2017kqp,Adam:2019srw} accounts only for
the kinematic vorticity associated with the velocity fields without
contribution from the temperature gradients and acceleration. It was confirmed by other calculations that the kinematic vorticity
alone describes the RHIC results much better than the thermal
vorticity~\cite{Florkowski:2019voj}. In addition, the
chiral kinetic approach with AMPT initial
conditions~\cite{Sun:2018bjl}, accounting for the transverse
vorticity fields due to deviation from longitudinal boost invariance,
generates the correct sign for $\pzstwo$. The difference in the sign of $\pzstwo$ between the experimental data and model calculations
based on solely thermal vorticity has been a subject of intense
investigations~\cite{Xia:2018tes,Sun:2018bjl, Florkowski:2019voj,
  Huang:2020dtn, Fu:2020oxj}.

Recently a possible explanation to the experimentally observed
positive $\pzstwo$ at RHIC was proposed based on the additional
contribution to polarization from fluid
shear~\cite{Liu:2021uhn,Becattini:2021suc}. The studies in
Refs.~\cite{Becattini:2021iol, Fu:2021pok} demonstrate that the fluid
shear competes with thermal vorticity and contributes with an opposite
phase to the azimuthal angle dependence of hyperon spin
polarization. Under the assumptions of isothermal (at constant
temperature) hadronization or that the hyperons
inherit the spin polarization of the constituent strange quark, the
effect of shear prevails over thermal vorticity and their combined
effect qualitatively explains the experimentally observed
azimuthal angle dependence of the hyperon spin polarization in Au--Au
collisions at $\sqrt{s_{\rm NN}} = 200$~GeV~\cite{Becattini:2021iol,
  Fu:2021pok}.  These studies indicate that the longitudinal
polarization is very sensitive to the hydrodynamic gradients and the
evolution of the spin degrees of freedom through different stages of
the evolution of the system created in heavy-ion collisions. The
measurement of $\pzstwo$ in Pb--Pb collisions at $\sqrt{s_{\rm NN}} =
5.02$~TeV and its comparison with measurements at RHIC as well as
theoretical models can provide important insights into the fluid
and spin dynamics in heavy--ion collisions.

In this Letter, we report the centrality, transverse momentum ($\pt$),
and rapidity ($y_{\rm H}$) dependences of $\pzstwo$ for $\Lambda$ ($\overline{\Lambda}$) hyperons measured by the
ALICE Collaboration in Pb--Pb collisions at $\sqrt{s_{\rm NN}} =
5.02$~TeV and compare with the previous STAR measurements in Au--Au
collisions at $\sqrt{s_{\rm NN}} = $ 200~GeV as well as with shear and
vorticity based hydrodynamic model calculations for Pb--Pb collisions
at $\sqrt{s_{\rm NN}} = 5.02$~TeV.

As the spin of a particle cannot be measured directly, the parity
violating weak decays of \mbox{$\Lambda\rightarrow {\rm p}+\pi^-$} and
$\overline\Lambda\rightarrow \overline {\rm p}+\pi^+$ in which the
momentum of the daughter (anti)proton is correlated with the spin of
the hyperon, are used to measure the polarization. The
angular distribution of the (anti)proton in the hyperon rest frame is
given by~\cite{Lee:1957qs}:
\begin{equation}
\begin{split} 
4\pi \frac{{\rm d}N}{{\rm d}\Omega^{*}} 
 = 1 + \alpha_{\rm H} {\bf P}_{\rm H} \cdot { \bf  \hat{p}}_{\rm p}^{*}                
 = 1 + \alpha_{\rm H}P_{\rm H}\cos\theta_{\rm p}^{*},
\label{Eq1}   
\end{split}                         
\end{equation}
where ${\bf P}_{\rm H}$ is the polarization vector, $\alpha_{\rm H}$
is the hyperon decay parameter ($\alpha_{\Lambda} = 0.750 \pm 0.009,
\alpha_{\overline{\Lambda}} = - 0.758 \pm 0.01$
~\cite{Ablikim:2018zay}), ${ \bf \hat{p}}_{\rm p}^{*}$ is the unit
vector along the (anti)proton momentum in the hyperon rest frame, and
$\theta_{\rm p}^{*}$ is the angle between the (anti)proton momentum
and the polarization vector in the hyperon rest frame. To measure the
polarization component along the $z$ direction, $\theta_{\rm p}^{*}$
is considered as the polar angle of the (anti)proton momentum in the
hyperon rest frame. The polarization $P_{\rm z}$ can be estimated by
averaging $\cos\theta_{\rm p}^{*}$ over all hyperons in all
collisions~\cite{Adam:2019srw}
\begin{equation}
P_{\rm z}(p_{\rm T}, y_{\rm H}, \varphi) 
= \frac{\mean{\cos\theta_{\rm p}^{*}}}
{\alpha_{\rm H}\mean{(\cos\theta_{\rm p}^{*})^{2}}},
\end{equation}
where $p_{\rm T}$, $y_{\rm H}$, and $\varphi$ are the transverse
momentum, rapidity, and azimuthal angle of the hyperon, respectively. The factor
$\mean{(\cos\theta_{\rm p}^{*})^{2}}$, which equals to $1/3$ in the case of
an ideal detector, is calculated directly from the data as a function
of centrality, $p_{\rm T}$, and $y_{\rm H}$ and serves as a correction
for finite acceptance along the longitudinal direction.

The data used in this analysis were collected by
ALICE~\cite{Aamodt:2008zz,Abelev:2014ffa} at the LHC in 2018 for
Pb--Pb collisions at $\sqrt{s_{\rm NN}} =$ 5.02 TeV. Two datasets,
corresponding to positive and negative magnetic field polarities, are
considered for this measurement. The centrality is determined using
the sum of the charge deposited in the V0A ($2.8<\eta< 5.1$) and the
V0C ($-3.7 <\eta< -1.7$) scintillator arrays, denoted as the V0M
centrality~\cite{Abbas:2013taa}. The event selection is based
on the trigger criteria and quality of the event vertex reconstruction
using the Time Projection Chamber (TPC)~\cite{Dellacasa:2000bm} and
the Inner Tracking System (ITS)~\cite{Dellacasa:1999kf}. Events that
pass central, semi-central, or minimum-bias
trigger criteria with a $z$-component of the reconstructed event vertex
($V_{\rm z})$ within $\pm10$~cm are selected. 
To suppress the pile-up of multiple collisions in the TPC drift
volume, events with a TPC multiplicity beyond 5 times the width of its
distribution at any V0M centrality are rejected. A similar cut on the
ITS centrality for the corresponding V0M centrality is applied to get
rid of additional outliers in the sample. In total about 270M events
are selected for the polarization measurement. The centrality
dependence of $\pzstwo$ is studied with 10\% centrality intervals
whereas the $p_{\rm T}$ and $y_{\rm H}$ dependence is studied in the
semi-central (30--50\%) collisions where elliptic flow is the largest.

The \lam and $\overline\Lambda$ hyperons are reconstructed inside the
TPC using the decay topology of \mbox{$\Lambda\rightarrow {\rm
    p}+\pi^-$} and $\overline\Lambda\rightarrow \overline {\rm
  p}+\pi^+$ (64\% branching ratio)~\cite{Zyla:2020zbs} as described in
Refs.~\cite{Abelev:2014ffa,Acharya:2018zuq}. The daughter tracks are
assigned the identity of a pion or a (anti)proton based on the charge
and particle identification using the specific energy loss (${\rm
  d}E/{\rm d}x$) measurement in the TPC. The tracks of the daughter
pions and (anti)protons are selected within the pseudorapidity range
of $|\eta| <0.8$ inside the TPC. Topological cuts such as distance of
closest approach (DCA) of the \lam and $\overline\Lambda$ candidates to
the primary vertex ($<$ 1.5 cm), DCA of the daughter tracks to the
primary vertex ($>$ 0.05 cm), DCA between the daughter tracks ($<$ 0.5
cm), and cosine of the pointing angle which is the angle between the
momentum direction of the hyperon and the direction from the primary
vertex to the decay point ($>$ 0.997) are used to reduce the
combinatorial background contribution to the invariant mass
spectrum. The \lam and $\overline\Lambda$ candidates having
\mbox{$1.103 < M_{\rm inv} < 1.129$~GeV/$c^{2}$} with \mbox{$\pt >
  0.5$}~GeV/$c$ and \mbox{$|y_{\rm H}|<0.5$} are considered in this
measurement.

The event-plane method is used for the polarization
measurement~\cite{Poskanzer:1998yz}. The second harmonic event plane
is reconstructed using the TPC tracks, and signals in the V0A and V0C
scintillators. The $X_{2}$ and $Y_{2}$ components of the second
harmonic flow vector are given by
\begin{equation}
 X_{2} 
= \frac{\sum_{\rm i}{w_{\rm i}\cos(2\varphi_{\rm i})}}{\sum{w_{\rm i}}},
  \quad
Y_{2}
= \frac{\sum_{\rm i}{w_{\rm i}\sin(2\varphi_{\rm i})}}{\sum{w_{\rm i}}},
\end{equation}
where, in case of the TPC, $w_{\rm i}=1$, $\varphi_{\rm i}$ is the
azimuthal angle of track $i$, and the sum runs over all the tracks
used in the flow vector construction. In the case of V0A and V0C, which
consist of four concentric rings with each ring divided into eight
segments, $\varphi_{\rm i}$ is the azimuthal angle of the centre of
the $i$-th segment and $w_{\rm i}$ is the measured signal proportional to the number of
particles detected in that segment.

\begin{figure}[htb]
\center
\includegraphics[width=0.50\textwidth]{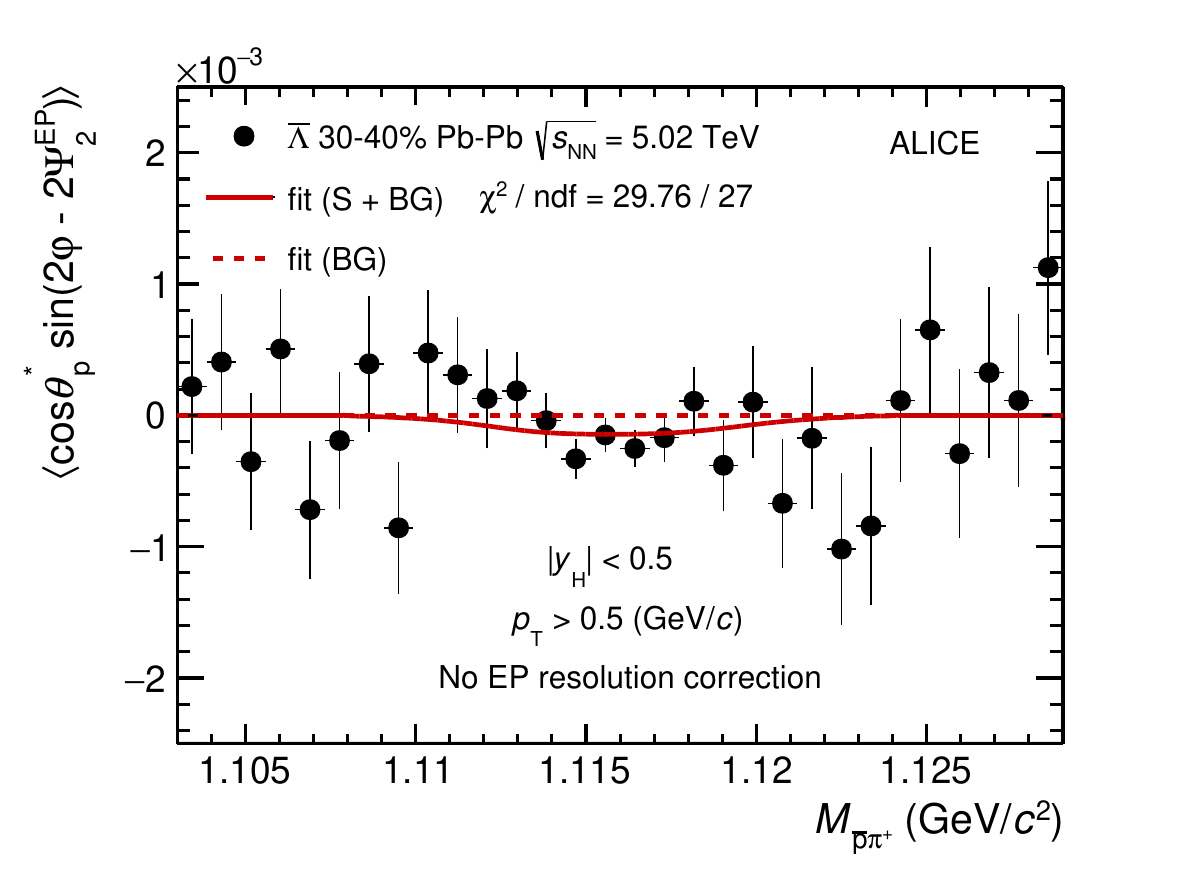}
\caption{ (color online) Fit to the invariant mass dependence of the
  $\mean{\cos\theta^*_{\rm p}  \sin(2\varphi - 2\psitwoep)}$ for
  $\overline\Lambda$ before event-plane resolution correction using Eq.~\ref{eq:Qfit} in the 30--40\% centrality class. See text for details. }
\label{invmass_fit_pol}
\end{figure}

The TPC flow vectors are reconstructed using tracks in the positive
($0.1 < \eta < 0.8$) and negative ($-0.8 < \eta < -0.1$) pseudorapidity regions and transverse momentum within $0.2 < \pt <3.0$~GeV/$c$.
The flow vector for the V0A or the V0C is constructed by averaging the
flow vectors of four rings using the energy deposited in each ring as a
weight. Because of the imperfect detector acceptance, varying beam
conditions, the averages $\mean{X_{\rm 2}}$ and $\mean{Y_{\rm
    2}}$ might deviate from zero.  To compensate these variations,
flow vectors are re-centered~\cite{Poskanzer:1998yz} run-by-run as a
function of the event centrality, event vertex position ($V_{\rm x}$,
$V_{\rm y}$, $V_{\rm z}$), and the time the event was taken during the run:
\begin{equation}
\label{eq:QvectorCorr}
  X_{\rm 2}' = X_{\rm 2} - \left<X_{\rm 2}\right>,
  \quad
  Y_{\rm 2}' = Y_{\rm 2} - \left<Y_{\rm 2}\right>.
\end{equation}
The second harmonic event-plane angle ($\psitwoep$) is constructed
from the re-centered flow vector components as
\begin{equation}
\label{eq:psi2}
\psitwoep = \frac{1}{2}\, {\rm tan^{\rm -1}}(Y_{\rm 2}'/X_{\rm 2}').
\end{equation}

The $\pzstwo$ is experimentally measured using $\psitwoep$ via
\begin{equation}
\pzstwo = \frac {\mean{{P_{\rm z} \sin(2\varphi - 2 \psitwoep)}}} {R(\psitwoep)},
\label{Eq110}   
\end{equation}
where $\varphi$ is the azimuthal emission angle of the particle, and
$\psitwoep$ is the reconstructed second harmonic event plane angle. The
$R(\psitwoep)$ is the event plane resolution correction~\cite{Poskanzer:1998yz}. 
The $\pzstwo$ is estimated using the invariant mass method~\cite{Adam:2019srw} by
calculating $Q = \mean{\cos\theta^*_{\rm p} \sin(2\varphi -
  2\psitwoep)}$ for all hyperon candidates as a function of the
invariant mass
and fitting it with the expression:
\begin{equation}
Q(M_{\rm inv})= f^{\rm S}(M_{\rm inv}) Q^{\rm S} +f ^{\rm BG} Q^{\rm BG}(M_{\rm inv}),
\label{eq:Qfit}
\end{equation}
where $f^{\rm S}$ and $f^{\rm BG}=1-f^{\rm S}$ are the signal and
background fraction, respectively, of the \lam and $\overline\Lambda$ candidates
estimated from the invariant mass yields. The constant $Q^{\rm S}$
estimates the signal and $Q^{\rm BG}(M_{\rm inv})$ estimates the
possible contribution from the combinatorial background of \lam and 
$\overline\Lambda$ hyperons towards the measured
polarization. By default, the results are obtained using the assumption of zero background contribution ($Q^{\rm BG}(M_{\rm inv}) = 0$) to the hyperon polarization. The results obtained with the assumption of $Q^{\rm BG}(M_{\rm inv})$ being a linear function of $M_{\rm inv}$ are found to be consistent with the default case indicating the measured polarization is not sensitive to this assumption. Figure~\ref{invmass_fit_pol} shows an example of the fit
to the invariant mass dependence of $Q(M_{\rm inv})$ using
Eq.~\ref{eq:Qfit} for $\overline\Lambda$ in the 30--40\% centrality
class. The $\pzstwo$ is obtained from $Q^{\rm S}$ after accounting
for the finite detector acceptance ($\mean{(\cos\theta^{*}_{\rm p})^{2}}$),
event-plane resolution correction, and scaling it with the hyperon decay constant
($\alpha_{\rm H}$).


The correction for the resolution of the second order event planes reconstructed in the
TPC, V0A, and V0C detectors are estimated using the three-subevent method
with the set of [TPC, V0A, V0C] and [V0, TPC-left ($-0.8<\eta<-0.1$),
TPC-right ($0.1<\eta< 0.8$)] event planes~\cite{Poskanzer:1998yz}. For
mid-central collisions, the event-plane resolution correction peaks at $\sim
0.88$ for the TPC and $\sim 0.84$ for the combined V0A and V0C
detectors. The results obtained using the event planes reconstructed
in the TPC and V0 detectors are found to be consistent with each
other and are combined to reduce the statistical uncertainty
considering the correlations between the event planes reconstructed in
two detectors. The $\pzstwo$ measured for \lam and $\overline\Lambda$
hyperons are consistent with each other as expected for the
polarization due to the elliptic flow induced vorticity and combined
to calculate the average hyperon polarization along the beam
direction. A large fraction of the measured \lam and
$\overline\Lambda$ hyperons originate from the decay of heavier
resonances. In Ref.~\cite{Becattini:2016gvu} it was shown that under
the assumption of similar vorticity induced polarization for all
final-state particles, the effect of feed-down is small, of the order
of 15\%. Similar to the previous STAR measurement~\cite{Adam:2019srw},
this measurement is not corrected for this effect.

The systematic uncertainties of this measurement are evaluated by
varying the criteria for the selection of the events, hyperon
daughters and topology of the decay, assumptions on the possible
contributions from the \lam and $\overline\Lambda$ background towards
the measured polarization, the $\pt$ dependent reconstruction
efficiency, and comparing results obtained with different magnetic
field orientations. The efficiency is estimated from
a Monte Carlo event generator HIJING~\cite{Wang:1991hta} by transporting the generated
particles through GEANT3~\cite{Brun:118715} simulated
detector response and performing track reconstruction in the ALICE
reconstruction framework. The effect
of the efficiency dependence on the hyperon transverse momentum is
found to be negligible. The differences between the results estimated
with the default and varied parameters, if found statistically
significant from the Barlow criterion~\cite{Barlow:2002yb}, are
considered as a source of systematic uncertainty. The Barlow criterion
is applied for each interval of centrality, $p_{\rm T}$, and $y_{\rm
  H}$ for which the final polarization results are presented. If the
Barlow criterion passes for more than 25\% of the total intervals, the
contribution of that particular systematic source is included in the
measurement uncertainty. The contributions from the different sources are added
in quadrature to estimate the total systematic uncertainty.

\begin{figure}[htb]
\center
\includegraphics[width=0.50\textwidth]{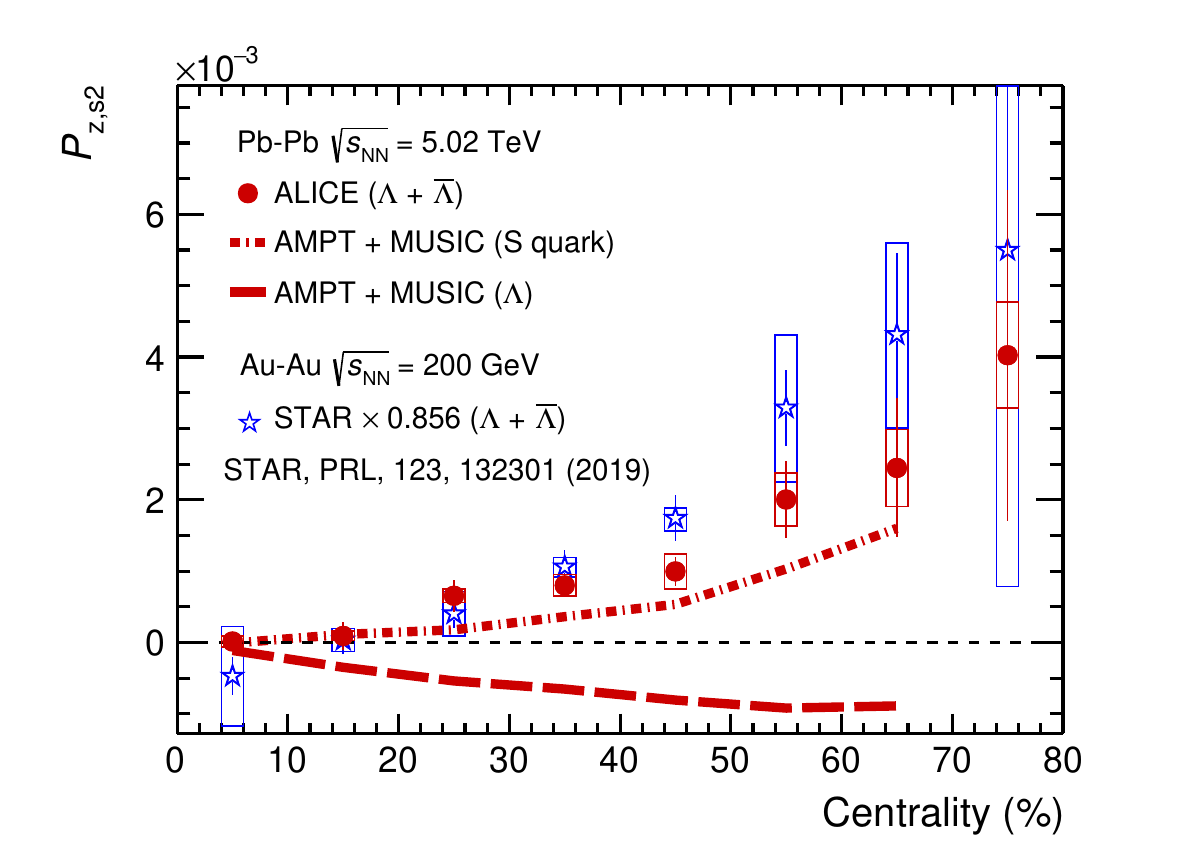}
\caption{ (color online) Centrality dependence of $\pzstwo$ averaged for \lam and
  $\overline\Lambda$ in Pb--Pb collisions at $\sqrt{s_{\rm NN}} =$
  5.02 TeV and its comparison with the RHIC results for Au--Au
  collisions at $\sqrt{s_{\rm NN}} =$ 200 GeV. The model
  calculations~\cite{hydroShear} for \lam and strange quark for Pb--Pb
  collisions at $\sqrt{s_{\rm NN}} =$ 5.02 TeV using the approach
  described in Ref.~\cite{Fu:2021pok} are shown by dash-dotted
  lines.  }
\label{lpol_star_alice_comp_cent}
\end{figure}
\begin{figure}[htb]
\center
\includegraphics[width=0.50\textwidth]{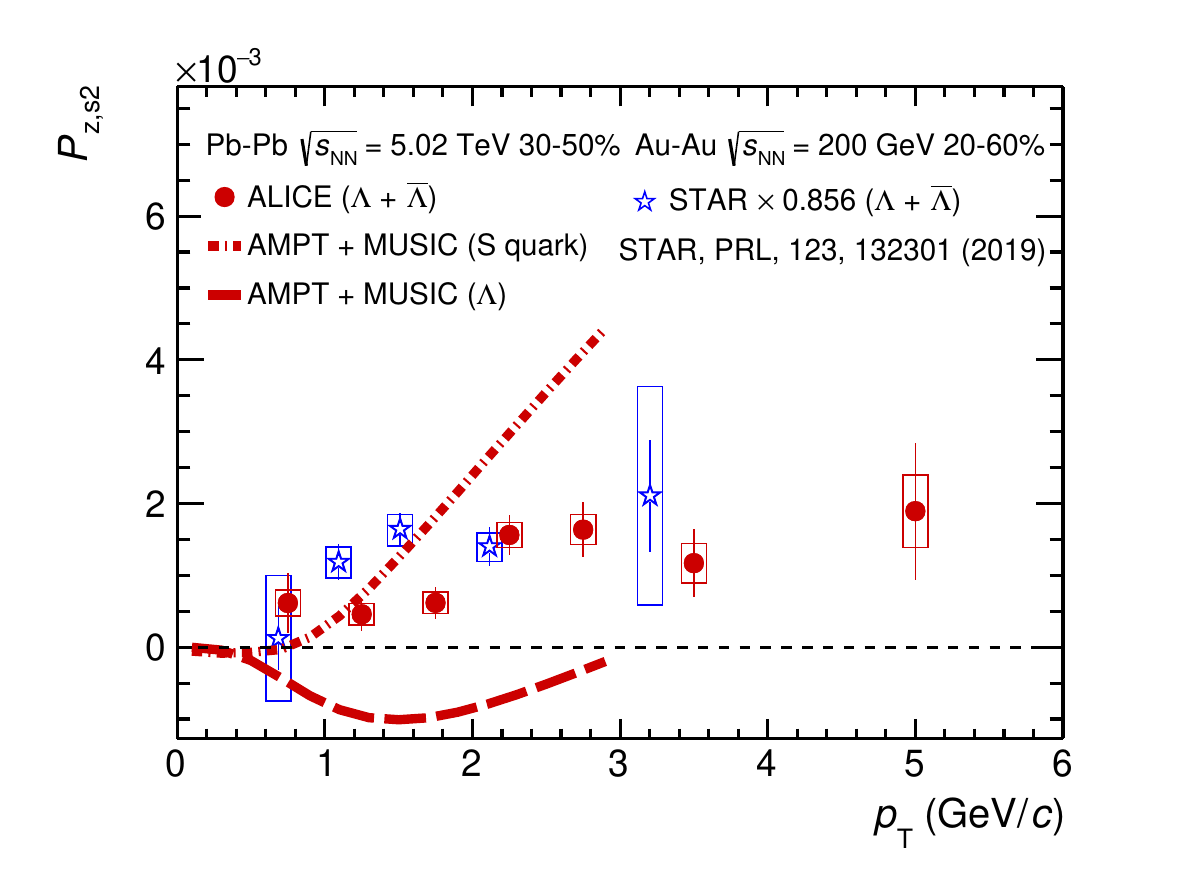}
\caption{ (color online) Transverse momentum dependence of
  $\pzstwo$ averaged for \lam and
  $\overline\Lambda$ in Pb--Pb collisions at $\sqrt{s_{\rm NN}} =$
  5.02 TeV in semi-central collisions and its comparison with the
  similar RHIC results for Au--Au collisions at $\sqrt{s_{\rm NN}} =$
  200 GeV.  The model calculations~\cite{hydroShear} for \lam and
  strange quark for Pb--Pb collisions at $\sqrt{s_{\rm NN}} =$ 5.02
  TeV in the 30--50\% centrality interval using the approach described
  in Ref.~\cite{Fu:2021pok} are shown by dash-dotted lines.  }
\label{lpol_star_alice_comp_pt}
\end{figure}
\begin{figure}[htb]
\center
\includegraphics[width=0.50\textwidth]{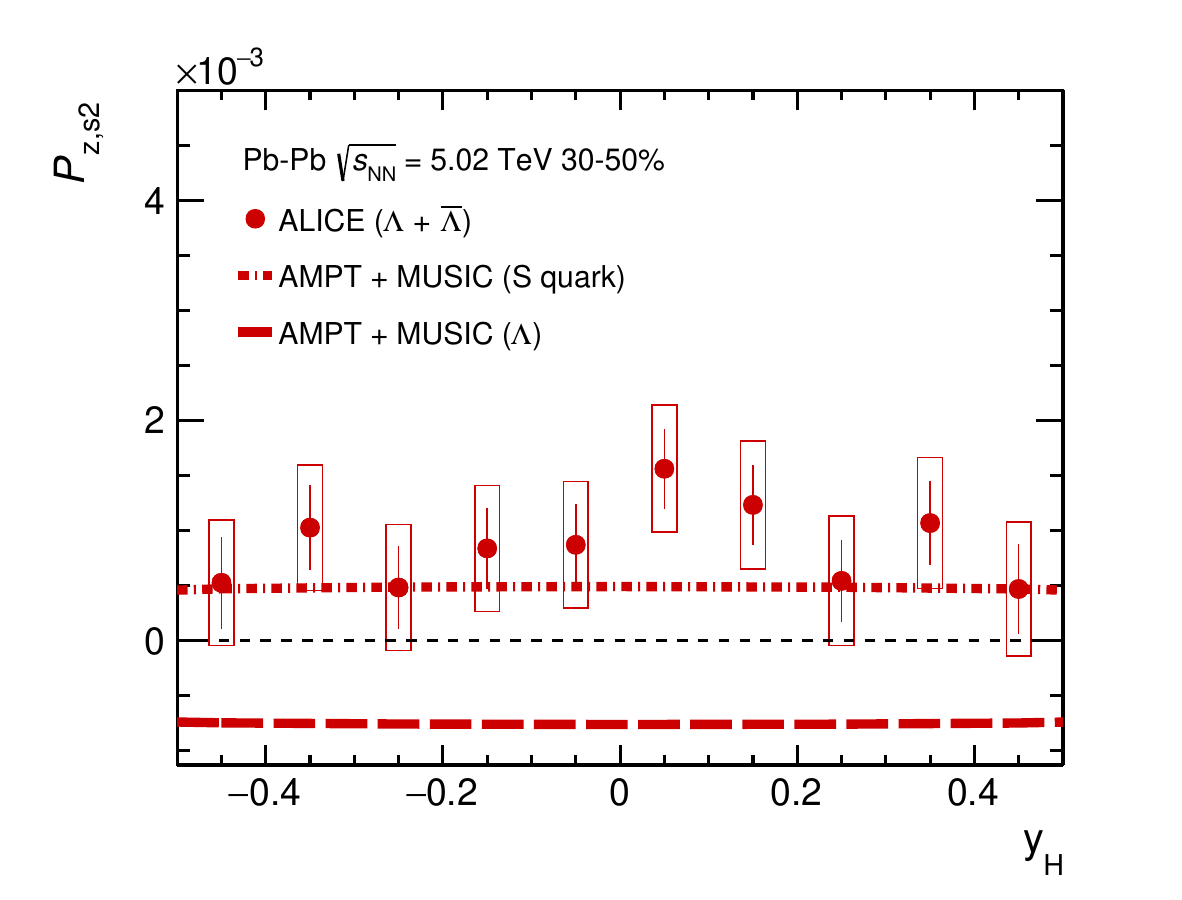}
\caption{ (color online) The rapidity dependence of $\pzstwo$ averaged for \lam and
  $\overline\Lambda$ in Pb--Pb collisions at $\sqrt{s_{\rm NN}} =$
  5.02 TeV in semi-central collisions. The model calculations~\cite{hydroShear} for \lam and
  strange quark for Pb--Pb collisions at $\sqrt{s_{\rm NN}} =$ 5.02 in
  the 30--50\% centrality interval using the approach described in
  Ref.~\cite{Fu:2021pok} are shown by dash-dotted lines.}
\label{lpol_rap}
\end{figure}

The centrality, $p_{\rm T}$, and $y_{\rm H}$ dependences of $\pzstwo$
in Pb--Pb collisions at $\sNN = 5.02$~TeV are shown in
Figs.~\ref{lpol_star_alice_comp_cent},~\ref{lpol_star_alice_comp_pt},
and~\ref{lpol_rap}. The $\pzstwo$ decreases towards more central
collisions, similar to the elliptic flow. For centralities larger than
60\%, the large uncertainties prevent a firm conclusion on its
centrality dependence. The $\pzstwo$ also shows an increase with
$p_{\rm T}$ up to $\pt \approx 2.0$~GeV/$c$ in the 30--50\% centrality
interval. For higher $p_{\rm T}$ ($p_{\rm T} > 2.0$~GeV/$c$), the
$\pzstwo$ is consistent with being constant but the uncertainty in the
measurement does not allow for a strong conclusion. The ALICE results
are compared with the STAR measurements in Au--Au collisions at $\sNN
= 200$~GeV~\cite{Adam:2019srw} in
Figs.~\ref{lpol_star_alice_comp_cent} and
\ref{lpol_star_alice_comp_pt}. As the STAR results were obtained with
$\alpha_{\rm H} =$ 0.642 whereas the ALICE measurement uses updated
values $\alpha_{\rm H} =$ 0.750 (\lam) and $-0.758$
($\overline\Lambda$), the STAR results are rescaled with a factor
0.856 for a proper comparison. Figure~\ref{lpol_star_alice_comp_cent}
indicates that the hyperon polarization in Pb--Pb collisions at $\sNN
= 5.02$~TeV is similar in magnitude for the central collisions with
somewhat smaller value in the semi-central collisions compared to the
top RHIC energy. The latter seems to originate at lower transverse
momenta ($p_{\rm T}<2.0$~GeV/$c$), where $\pzstwo$ at the LHC is
smaller than that at the top RHIC energy in semi-central collisions as
shown in Fig.~\ref{lpol_star_alice_comp_pt}. The $\pzstwo$ does not
exhibit a significant dependence on rapidity as shown in
Fig.~\ref{lpol_rap}.

The comparison between the ALICE results and the $\pzstwo$ values estimated from the fluid shear and thermal
vorticity in a hydrodynamic model following the scheme used
in Ref.~\cite{Fu:2021pok} is shown in Figs.~\ref{lpol_star_alice_comp_cent},~\ref{lpol_star_alice_comp_pt},
and~\ref{lpol_rap}. The 3+1 D hydrodynamical model
MUSIC~\cite{Schenke:2010nt,Schenke:2010rr} with AMPT initial
conditions~\cite{Lin:2004en,Xu:2016hmp}, tuned to describe the $dN_{\rm ch}/d\eta$~\cite{Adam:2015ptt}, $p_{\rm T}$
spectra~\cite{Acharya:2019yoi}, and $v_{2}(p_{\rm T})$ of pions, kaons,
and protons~\cite{Acharya:2018zuq,Zhao:2017yhj} in Pb--Pb collisions
at $\sNN =5.02$~TeV is used for the longitudinal polarization
calculation. In the
scenario where the polarization is calculated for the
\lam and $\overline\Lambda$ at the freeze-out using hyperon mass for the
mass of the spin carrier, the effect of thermal vorticity dominates
over the shear induced polarization and total $\pzstwo$ shows a negative sign. However,
considering the constituent strange quark as the spin carrier and the
hyperons inheriting the spin polarization of the strange quark at
hadronization, the effect of fluid shear prevails over thermal vorticity and generates
the correct sign for resulting $\pzstwo$ as shown in
Figs.~\ref{lpol_star_alice_comp_cent},~\ref{lpol_star_alice_comp_pt},
and~\ref{lpol_rap}.  In both cases, the effect of hadronic
scatterings on the hyperon spin polarization is not considered. Note that theoretical models, 
including the spin degrees of freedom
consistently through all the stages of heavy-ion collisions are not
yet well developed~\cite{Sheng:2020ghv,Wang:2021owk}. Comparison of
the experimental results with the two scenarios discussed
here provides a qualitative idea about the possible consequences of
the different assumptions used to estimate hyperon polarization from
the velocity and temperature gradients generated by hydrodynamic or transport
models~\cite{Fu:2021pok}.


In summary, the polarization of \lam and $\overline\Lambda$
hyperons along the beam direction has been measured in Pb--Pb
collisions at $\sNN = 5.02$~TeV using the ALICE detector at the
LHC. The polarization exhibits a clear second harmonic sine modulation
as expected due to elliptic flow. This is the first experimental
evidence of a significant $z$-component of the hyperon polarization
due to elliptic flow induced vorticity at the LHC. The $\pzstwo$
measured in Pb--Pb collisions at $\sNN = 5.02$~TeV is of similar
magnitude as the one measured by the STAR Collaboration in Au--Au
collisions at $\sNN = 200$~GeV. 
No significant
dependence of $\pzstwo$ on the rapidity is observed. The sign of the
$\pzstwo$ is positive at both RHIC and the LHC and in disagreement
with hydrodynamic and AMPT models estimations accounting only for the
thermal vorticity. The introduction of shear induced
polarization~\cite{Becattini:2021iol, Fu:2021pok} along with
additional assumptions on the hadronization temperature or mass of the
spin carrier reproduces the experimentally observed positive $\pzstwo$ at RHIC
and the LHC energies. These studies indicate that longitudinal polarization is sensitive to the hydrodynamic gradients as well as the dynamics of the spin degrees of freedom through the different stages of the evolution of the system created in heavy-ion collisions. For a quantitative data to model comparison, a detailed theoretical understanding of the quark spin polarization in the quark--gluon plasma (QGP), spin transfer at the hadronization, and the effect of hadronic scattering on the spin polarization are required. The upcoming Run 3 at the LHC will provide much larger data samples for more differential and precision measurements of local and global hyperon polarization and provide further constraints to the models aiming to explain the vorticity and the particle polarization in heavy-ion collisions.

%


\newenvironment{acknowledgement}{\relax}{\relax}
\begin{acknowledgement}
\section*{Acknowledgements}

The ALICE Collaboration would like to thank all its engineers and technicians for their invaluable contributions to the construction of the experiment and the CERN accelerator teams for the outstanding performance of the LHC complex.
The ALICE Collaboration gratefully acknowledges the resources and support provided by all Grid centres and the Worldwide LHC Computing Grid (WLCG) collaboration.
The ALICE Collaboration acknowledges the following funding agencies for their support in building and running the ALICE detector:
A. I. Alikhanyan National Science Laboratory (Yerevan Physics Institute) Foundation (ANSL), State Committee of Science and World Federation of Scientists (WFS), Armenia;
Austrian Academy of Sciences, Austrian Science Fund (FWF): [M 2467-N36] and Nationalstiftung f\"{u}r Forschung, Technologie und Entwicklung, Austria;
Ministry of Communications and High Technologies, National Nuclear Research Center, Azerbaijan;
Conselho Nacional de Desenvolvimento Cient\'{\i}fico e Tecnol\'{o}gico (CNPq), Financiadora de Estudos e Projetos (Finep), Funda\c{c}\~{a}o de Amparo \`{a} Pesquisa do Estado de S\~{a}o Paulo (FAPESP) and Universidade Federal do Rio Grande do Sul (UFRGS), Brazil;
Ministry of Education of China (MOEC) , Ministry of Science \& Technology of China (MSTC) and National Natural Science Foundation of China (NSFC), China;
Ministry of Science and Education and Croatian Science Foundation, Croatia;
Centro de Aplicaciones Tecnol\'{o}gicas y Desarrollo Nuclear (CEADEN), Cubaenerg\'{\i}a, Cuba;
Ministry of Education, Youth and Sports of the Czech Republic, Czech Republic;
The Danish Council for Independent Research | Natural Sciences, the VILLUM FONDEN and Danish National Research Foundation (DNRF), Denmark;
Helsinki Institute of Physics (HIP), Finland;
Commissariat \`{a} l'Energie Atomique (CEA) and Institut National de Physique Nucl\'{e}aire et de Physique des Particules (IN2P3) and Centre National de la Recherche Scientifique (CNRS), France;
Bundesministerium f\"{u}r Bildung und Forschung (BMBF) and GSI Helmholtzzentrum f\"{u}r Schwerionenforschung GmbH, Germany;
General Secretariat for Research and Technology, Ministry of Education, Research and Religions, Greece;
National Research, Development and Innovation Office, Hungary;
Department of Atomic Energy Government of India (DAE), Department of Science and Technology, Government of India (DST), University Grants Commission, Government of India (UGC) and Council of Scientific and Industrial Research (CSIR), India;
Indonesian Institute of Science, Indonesia;
Istituto Nazionale di Fisica Nucleare (INFN), Italy;
Institute for Innovative Science and Technology , Nagasaki Institute of Applied Science (IIST), Japanese Ministry of Education, Culture, Sports, Science and Technology (MEXT) and Japan Society for the Promotion of Science (JSPS) KAKENHI, Japan;
Consejo Nacional de Ciencia (CONACYT) y Tecnolog\'{i}a, through Fondo de Cooperaci\'{o}n Internacional en Ciencia y Tecnolog\'{i}a (FONCICYT) and Direcci\'{o}n General de Asuntos del Personal Academico (DGAPA), Mexico;
Nederlandse Organisatie voor Wetenschappelijk Onderzoek (NWO), Netherlands;
The Research Council of Norway, Norway;
Commission on Science and Technology for Sustainable Development in the South (COMSATS), Pakistan;
Pontificia Universidad Cat\'{o}lica del Per\'{u}, Peru;
Ministry of Education and Science, National Science Centre and WUT ID-UB, Poland;
Korea Institute of Science and Technology Information and National Research Foundation of Korea (NRF), Republic of Korea;
Ministry of Education and Scientific Research, Institute of Atomic Physics and Ministry of Research and Innovation and Institute of Atomic Physics, Romania;
Joint Institute for Nuclear Research (JINR), Ministry of Education and Science of the Russian Federation, National Research Centre Kurchatov Institute, Russian Science Foundation and Russian Foundation for Basic Research, Russia;
Ministry of Education, Science, Research and Sport of the Slovak Republic, Slovakia;
National Research Foundation of South Africa, South Africa;
Swedish Research Council (VR) and Knut \& Alice Wallenberg Foundation (KAW), Sweden;
European Organization for Nuclear Research, Switzerland;
Suranaree University of Technology (SUT), National Science and Technology Development Agency (NSDTA) and Office of the Higher Education Commission under NRU project of Thailand, Thailand;
Turkish Energy, Nuclear and Mineral Research Agency (TENMAK), Turkey;
National Academy of  Sciences of Ukraine, Ukraine;
Science and Technology Facilities Council (STFC), United Kingdom;
National Science Foundation of the United States of America (NSF) and United States Department of Energy, Office of Nuclear Physics (DOE NP), United States of America.
\end{acknowledgement}

\bibliographystyle{utphys}   
\bibliography{bibilography}

\newpage
\appendix

%
%

\section{The ALICE Collaboration}
\label{app:collab}

\begin{flushleft}

S.~Acharya$^{\rm 143}$, 
D.~Adamov\'{a}$^{\rm 98}$, 
A.~Adler$^{\rm 76}$, 
G.~Aglieri Rinella$^{\rm 35}$, 
M.~Agnello$^{\rm 31}$, 
N.~Agrawal$^{\rm 55}$, 
Z.~Ahammed$^{\rm 143}$, 
S.~Ahmad$^{\rm 16}$, 
S.U.~Ahn$^{\rm 78}$, 
I.~Ahuja$^{\rm 39}$
% ...

\bigskip

\bigskip 

\textbf{\Large Affiliation Notes}

\bigskip 

% ...
$^{\rm VII}$ Also at: University of Kansas, Lawrence, Kansas, United States\\

\bigskip

\bigskip 

\textbf{\Large Collaboration Institutes}

\bigskip 

% ...
$^{16}$ Department of Physics, Aligarh Muslim University, Aligarh, India\\
% ...
$^{31}$ Dipartimento DISAT del Politecnico and Sezione INFN, Turin, Italy\\
% ...
$^{35}$ European Organization for Nuclear Research (CERN), Geneva, Switzerland\\
% ...
$^{39}$ Faculty of Science, P.J.~\v{S}af\'{a}rik University, Ko\v{s}ice, Slovakia\\
% ...
$^{55}$ INFN, Sezione di Bologna, Bologna, Italy\\
% ...
$^{76}$ Johann-Wolfgang-Goethe Universit\"{a}t Frankfurt Institut f\"{u}r Informatik, Fachbereich Informatik und Mathematik, Frankfurt, Germany\\
% ...
$^{78}$ Korea Institute of Science and Technology Information, Daejeon, Republic of Korea\\
% ...
$^{98}$ Nuclear Physics Institute of the Czech Academy of Sciences, \v{R}e\v{z} u Prahy, Czech Republic\\
% ...
$^{143}$ Variable Energy Cyclotron Centre, Homi Bhabha National Institute, Kolkata, India\\
% ...

\bigskip 

\end{flushleft} 
  
\end{document}